# Energy Efficient Power Allocation in Massive MIMO Systems with Power Limited Users

Abdolrasoul Sakhaei Gharagezlou [1], Mahdi Nangir [2], Nima Imani [3], Erfan Mirhosseini [4]

[1,2,3] Electrical and Computer Engineering Department, University of Tabriz, Tabriz, Iran
[4] Electrical Engineering Department, Amirkabir University of Technology, Tehran, Iran
[1] abdolrasoulsakhaei@gmail.com, [2] nangir@tabrizu.ac.ir, [3] nimaimani91@gmail.com, [4] erfanm@aut.ac.ir

*Abstract*— In this paper, we investigate energy-efficient (EE) power allocation (PA) for a special downlink scenario of the massive multiple-input multiple-output (MIMO) systems. We consider a minimum power required for each user to ensure that the quality of service (QoS) for each user is satisfied. In this method, a comparison between the sum of minimum power required by users and the maximum transmission power is done to determine whether maximizing EE is possible or not. If the sum of the minimum power required by users is less than the maximum transmission power, we maximize EE. Otherwise, the number of admitted users in a cluster is maximized. In both cases, the simulation results show that the proposed algorithm has better performance than similar algorithms.

*Keywords*— Massive multiple-input multiple-output, Energy efficiency, Power allocation, Power limited user, Admitted user.

## 1. INTRODUCTION

The MIMO systems, in which a base station (BS) is equipped with hundreds of antennas and communicates with tens or hundreds of users simultaneously in a same frequency-time block, are known as massive MIMO systems. As the number of BS antennas in massive MIMO systems goes to infinite, the noise and dimming effects disappear in the small scale range. Massive MIMO systems provide reliable communication, high energy efficiency and low-complexity schemes for signal processing. Several technologies are involved in the massive MIMO in fifth generation (5G) networks including non-orthogonal multiple access (NOMA), heterogeneous networks, millimeter waves and device-to-device communications [1].

The total sum-rate obtained by the power allocation method in multi-cell massive MIMO systems is studied in [2]. A new power allocation algorithm is proposed to increase the energy efficiency and capacity of massive MIMO systems [3]. A two-step iterative algorithm with a combination of antennas and users is proposed to maximize energy efficiency in [4]. The results presented in this paper show that energy efficiency with the maximum-ratio combining (MRC) receiver has been improved by 71.16% when the number of users equals 60. In [5], the authors examine allocation of the energy efficient power for the massive MIMO system with maximum ratio transmission (MRT) pre-coding scheme [5]. The problem of specialized power optimization in radio cognitive networks using the NOMA method is investigated [6]. In [7], to solve the power allocation problem for a MIMO-NOMA system in a cluster, a minimum rate is considered for each user. An optimal power allocation solution that can improve the total data rate of a mobile network cell with a reduced complexity is proposed in [8]. In [9], the minimum transmission power of each user is calculated using the maximal signal to leakage ratio (maximal-SLR) solution for each user. Furthermore, a new radio access scheme is proposed which consists of the relay protocol and the NOMA method [10]. A power allocation scheme is obtained for a set of parallel channels while the transmitter has partial channel status information [11]. The energy efficiency of a heterogeneous cellular network with massive MIMOs and Small Cells (SCs) is discussed in [12]. Furthermore, the energy-efficient power allocation for a multi-user massive MIMO system in a single cell is investigated by using the SIF method in [13].

In this paper, we propose an energy efficient power allocation scheme in the massive MIMO scenario which is motivated by [13]. The novelty and innovation of this paper can be summarized as follows.

In this work, a minimum power required by users is considered to provide quality of user service. The Optimization problem in this paper is divided into two parts. They are maximizing energy efficiency and maximizing the number of admitted users in a cluster. First, we determine which of the optimization problems should be addressed by comparing the sum of the minimum power required by users to ensure the quality of service with the maximum transmit power. If sum of the minimum power required by users is less than the maximum power output, the energy efficiency is maximized. Otherwise, the number of accepted users in a cluster is maximized. We propose a new iterative algorithm for the power allocation and consider a required power for satisfying the QoS of users in a cluster. It is shown that the proposed algorithm performs better than other similar power allocation algorithms.

The following four parts of this article are discussed below. In Section 2, the model of the system studied is described. In Section 3, the optimization problem is formulated and we obtain the optimal power allocation scheme by solving it. In Section 4, simulation results and performance analysis are presented. Finally, we conclude paper in Section 5.

## 2. SYSTEM MODEL

In this paper, we consider the system depicted in Fig. 1. In this system, the BS is equipped with $M$ antennas and each user is equipped with single antenna.

Let consider $\mathbf{G}$ is the matrix of the flat fading channel between the BS and $K$ users. $\mathbf{G}$ can be formulated as:

$$\mathbf{G} = \mathbf{H}\mathbf{D}^{1/2}. \qquad (1)$$

where, $\mathbf{H} \in \mathcal{C}^{M \times K}$ is the small scale fading channel matrix and $\mathbf{D} = \text{diag}\{\beta_1, \beta_2, \ldots, \beta_K\}$ denotes the large scale fading matrix, where $\beta_K = \varphi \varrho / d_k^\varepsilon$ represents the path loss and shadow fading. Parameter $\varphi$ is a constant related to the carrier frequency and antenna gain and $d_k$ is the distance between the BS and the $k$-th user. Furthermore, $\varepsilon$ represents the path loss exponent, and $\varrho$ is the shadow fading with lognormal distribution, i.e., $10\log_{10}(\varrho) \sim N(0.\sigma^2)$.

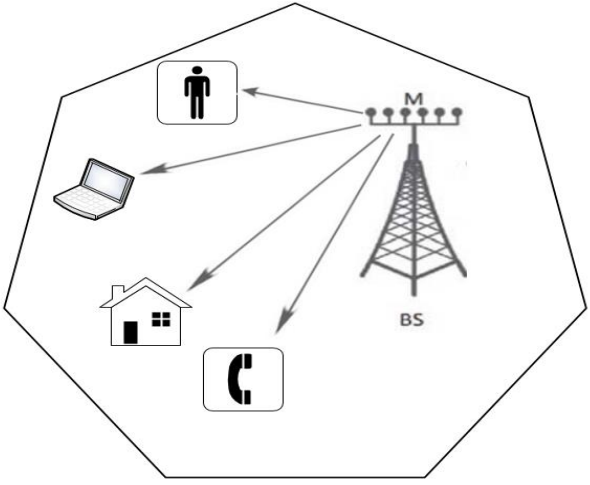

**Fig.1.** System model of the massive MIMO systems.

Accordingly, the observed signal of the $k$-th user is formulated as follows [13]:

$$y_k = \sqrt{p_k \beta_k}\|\mathbf{h}_k\|s_k + \sum_{l=1.l\neq k}^{K}\sqrt{p_l\beta_k}\frac{\mathbf{h}_k\mathbf{h}_l^H}{\|\mathbf{h}_l\|} + n_k. \quad (2)$$

where $k \in \{1.2.\dots.K\}$, $p_k$ denotes the transmit power allocated to the $k$-th user and $\mathbf{h}_k$ is the $k$-th column of $\mathbf{H}$. Besides, $s_k$ represents transmit data symbol of the $k$-th user and $n_k$ is the Additive White Gaussian Noise (AWGN) at the $k$-th user with distribution $N(0.N_0)$, where $N_0$ denotes the noise power spectral density.

The received SINR for $k$-th user can be expressed as:

$$\gamma_k = \frac{p_k\beta_k\|\mathbf{h}_k\|^2}{\left\|\mathbf{h}_k\sum_{l=1.\neq k}^{K}\sqrt{p_l\beta_k}\frac{\mathbf{h}_l^H}{\|\mathbf{h}_l\|}\right\|^2 + \sigma^2}. \quad (3)$$

where $\|.\|$ represents the $L2$-norm. The spectral efficiency for $k$-th user can be written as:

$$r_k = B\log_2\left(1 + \frac{\gamma_k}{\mu}\right). \quad (4)$$

where $\mu = \frac{2}{3}\ln(5P_e)$ is the SNR gap between the Shannon channel capacity and a practical modulation with coding scheme which achieves the BER value of $P_e$ [14]. Then, the total spectral efficiency achieved by all users can be expressed as:

$$r(\mathbf{P}) = \sum_{k=1}^{K} r_k = B\sum_{k=1}^{K}\log_2(1 + \frac{\gamma_k}{\mu}). \quad (5)$$

where $\mathbf{P} = [p_1.p_2.\dots.p_k]^T$ is the power allocation vector for all users from the BS.

According to description of (1), $\mathbf{h}_l$ ($l = 1.2.\dots.k-1.k+1.\dots.K$) are independent of random vector $\mathbf{h}_k$. If we define $\alpha_l \cong \left\|\frac{\mathbf{h}_k\mathbf{h}_l^H}{\|\mathbf{h}_l\|}\right\|^2$, then $\alpha_l$ will be a gamma random variable with parameters (1,1) and $\alpha_0 \cong \|\mathbf{h}_k\|^2$ is a gamma random variable with parameters $(M,1)$. Thus we have,

$$E\left[\left\|\mathbf{h}_k\sum_{l=1.\neq k}^{K}\sqrt{p_l\beta_k}\frac{\mathbf{h}_l^H}{\|\mathbf{h}_l\|}\right\|^2\right] =$$
$$\beta_k\sum_{l=1.\neq k}^{K}\sqrt{p_l}\,E[\alpha_l] = \beta_k\sum_{l=1.\neq k}^{K}\sqrt{p_l}. \quad (6)$$

and

$$E[p_k\beta_k\|\mathbf{h}_k\|^2] = Mp_k\beta_k. \quad (7)$$

Using (6) and (7), the equation (3) can be rewritten as follows [5]:

$$\gamma_k = \frac{p_k\beta_k M}{\beta_k\sum_{l=1.\neq k}^{K}\sqrt{p_l} + \sigma^2}. \quad (8)$$

In this work, the energy efficiency is defined as follows:

$$\eta_{EE} = \frac{\sum_{k=1}^{K} r_k}{\sum_{k=1}^{K} p_k + \sum_{m=1}^{M} P_{c.m}}. \quad (9)$$

where $P_{c.m}$ is the fixed circuit power consumption per antenna.

Our goal is to maximize the energy efficiency when each user has a pre-defined minimum required power. The optimization problem is formulated as follows:

$$maximize_{\{p_1.p_2\dots p_K\}}\ \eta_{EE} \quad (10.a)$$

$$C1: \sum_{k=1}^{K} p_k = P_T \quad (10.b)$$

$$C2: p_k \geq (\omega_k - 1)\left(\sum_{j=1}^{k-1} p_j + \frac{1}{\vartheta\|\mathbf{h}_k\|^2}\right). \quad (10.c)$$

where (10.b) shows the total BS power constraint and $P_T$ denotes the flexible transmit power. The constraints of (10.c) show QoS of users, i.e., the minimum required power, is guaranteed. In $C_2$, $\omega_k = 2^{R_k^{\text{Min}}}$ where $R_k^{\text{Min}}$ is the minimum required data rate for $k$-th user for $k = 1.2.\dots.K$. Furthermore, $\vartheta$ denotes the signal-to-noise ratio (SNR).

### 3. PROPOSED SOLUTION

Given the minimum power for each user (10.c), the optimization problem (10) may not a solution if the total transmission power is not large enough. If the problem of maximizing energy efficiency is not feasible, then we maximize the number of users in the cluster and hereto we provide an efficiency.

We compare sum of the minimum power required by users with the total transmission power for determining the feasibility of the optimization problem (10). The minimum power required by each user is equal to:

$$P_{req.k} = (\omega_k - 1)\left(\sum_{j=1}^{k-1} p_j + \frac{1}{\vartheta\|\mathbf{h}_k\|^2}\right). \quad (11)$$

which is the minimum required power to satisfy the QoS requirement the of $k$-th user. As a result, if

$$\sum_{k=1}^{K} P_{req.k} \leq P_{max}. \quad (12)$$

then the problem (10) is feasible; otherwise, it is not.

## A. EE maximization when problem (10) is feasible

The problem of maximizing energy efficiency can be formulated as:

$$maximize_{\{p_1,p_2,...,p_K\}} \quad \eta = \frac{R^{sum}}{P_c + \sum_{k=1}^{K} p_k} \quad (13.a)$$

$$C1: \sum_{k=1}^{K} p_k = P_T \quad (13.b)$$

$$C2: p_k \geq (\omega_k - 1)\left(\sum_{j=1}^{k-1} p_j + \frac{1}{\vartheta \|h_k\|^2}\right). \quad (13.c)$$

The objective function of (13) has a fraction form and it is a non-convex function. Using fractional programming and low data rate, problem (13) can be turned into a convex optimization problem. The maximum EE can be achieved if and only if,

$$maximize_{\{p_1,p_2,...,p_K\}}\left\{\sum_{k=1}^{K} r_k - q^*\left(\sum_{k=1}^{K} p_k + \sum_{m=1}^{M} P_{c,m}\right)\right\} = 0. \quad (14)$$

where $q^*$ represents the maximum EE [15]. By considering perfect channel state information (CSI), the Rayleigh fading and the MRT pre-coding, the lower bound of the data rate can be written as follows [5],

$$\tilde{r}_k = B\log_2\left(\frac{M\beta_k p_k}{\beta_k \sum_{\substack{j=1 \\ j \neq k}}^{K} p_j + \sigma^2}\right). \quad (15)$$

According to (15), the optimization problem in (13) can be converted to a simpler form as:

$$maximize_{\{p_1,p_2,...,p_K\}}\left\{\sum_{k=1}^{K} \tilde{r}_k - q^*\left(\sum_{k=1}^{K} p_k + \sum_{m=1}^{M} P_{c,m}\right)\right\} \quad (16.a)$$

$$C1: \sum_{k=1}^{K} p_k = P_T \quad (16.b)$$

$$C2: p_k \geq (\omega_k - 1)\left(\sum_{j=1}^{k-1} p_j + \frac{1}{\vartheta \|h_k\|^2}\right). \quad (16.c)$$

The simplified optimization problem in (16) is a constrained problem, hence the Lagrange function is used to convert it to a non-constrained problem [16]. Let,

$$\Phi(\mathbf{p},\theta,\lambda) = -\left[\sum_{k=1}^{K} r_k - q\left(\sum_{k=1}^{K} p_k + \sum_{m=1}^{M} P_{c,m}\right)\right]$$

$$-\lambda_k\left(p_k - (\omega_k - 1)\left(\sum_{j=1}^{k-1} p_j + \frac{1}{\vartheta \|h_k\|^2}\right)\right) \quad (17)$$

$$-\theta(P_T - \sum_{k=1}^{K} p_k).$$

where, **p** represents feasible set of power variables, $\theta \geq 0$ is the Lagrangian multiplier corresponding to the transmission power constraint. Moreover, $\lambda$ is the Lagrangian multiplier vector corresponding with the QoS of users (minimum required power), such that $\lambda_k \geq 0$.

Necessary and sufficient condition to obtain the optimal transmission power is expressed as:

$$\frac{\partial \Phi}{\partial p_k} = \sum_{j=1, j \neq k} \frac{1}{(\sum_{i=1 \neq k}^{K} p_i + \sigma^2/\beta_i)\ln 2} - \frac{1}{p_k \ln 2} \quad (18)$$

$$+ \left(\lambda_k - \sum_{j=1}^{k-1}(\omega_j - 1)\lambda_j\right) + \theta + q = 0.$$

Thus by using (18), the optimal power of each user is formulated as:

$$p_k = \frac{1}{\left(\sum_{j=1, j \neq k} \frac{1}{(\sum_{i=1 \neq k}^{K} p_i + \sigma^2/\beta_i)\ln 2} + \theta + q + \chi\right)\ln 2}. \quad (19)$$

where,

$$\chi = \left(\lambda_k - \sum_{j=1}^{k-1}(\omega_j - 1)\lambda_j\right). \quad (20)$$

## B. User admission when problem (10) is infeasible

If (10) is infeasible, then the problem of maximizing admitted users is interested and it is formulated as follows:

$$maximize_{\{p_1,p_2,...,p_K\}} \sum_{k=1}^{K} X_k \quad (21.a)$$

$$C1: \sum_{k=1}^{K} p_k = P_T \quad (21.b)$$

$$C2: p_k \geq (\omega_k - 1)\left(\sum_{j=1}^{k-1} p_j + \frac{1}{\vartheta \|h_k\|^2}\right) \quad (21.c)$$

$$C3: X_k \in \{0,1\}. \quad (21.d)$$

where $X_k$ is the binary decision variable indicating whether $k$-th user is admitted or not.

The user admission method is performed according to an iterative action. During each iteration, the user with the best channel gain is first selected. Among the selected users, their required power is calculated based on the interference of other accepted users. In the next step, this obtained power is compared with the total remaining power. If the total remaining power is higher than it, then this user is selected for acceptance in the cluster and it is removed from the candidates. Eventually, the remaining total power is updated. This process continues until no user meets the acceptance requirements for admission in the cluster [7]. The total remaining power is formulated as follows:

$$P_{remain} = P_{max} - P_{req}. \quad (22)$$

## 4. SIMULATION RESULTS

In this section, simulation results are presented to verify the performance of the proposed PA strategy and user admission scheme. The parameters used in our simulations are given in Table 1.

**Table 1.** Simulation Parameters

| Parameters | Value |
|---|---|
| RB bandwidth $B$ | 120 kHz |
| Number of transmit antennas $M$ | 128 |
| Number of users $K$ | 3 |
| Noise spectral density $N_0$ | -170 dBm/Hz |
| Variance of log-normal shadow fading $\sigma^2$ | 10 dB |
| Factor $\varphi$ | 1 |

In the following, we provide our proposed algorithm in a step by step procedure.

---

**Algorithm 1** Energy efficient power allocation algorithm for the massive MIMO system.

**Initialize 1**: $R_k^{Min}$ . $(k = 1.2. \dots K)$

If_1 ($\sum_{k=1}^{K} p_{k.min} \leq P_{max}$)

**Step 1**: **Initialize 2**: $u = 0$, $v = \frac{1}{\mu\sigma^2 \ln 2} \min_{k} \{\beta_k \|\mathbf{h}_k\|^2\}$, $\varepsilon > 0$

$p_k > 0$ $(k = 1.2 \dots K)$ transmit power and Lagrangian multipliers $P_T$, $\lambda^{(0)}, \theta^{(0)}$

**Step 2**: Let $\eta_{EE} = \frac{u+v}{2}$, solve $p_k$ according to formula (19),

  Then a realistic EE **q** can be got by formula (6),

  If_2 ($\eta_{EE} \geq \eta_{EE}^R$), then $u = \eta_{EE}$

  Else $v = \eta_{EE}$.

  End if_2

**Step 3: Update**

$$\theta^{(n+1)} = max\left(0. \theta^{(n)} - \varpi_1 \left(P_T - \sum_{k=1}^{K} p_k^{(n)}\right)\right)$$

$$\lambda^{(n+1)} = max(0. \lambda^{(n)} - \varpi_k(p_k - (\omega_k - 1)\left(\sum_{j=1, j \neq k}^{K} p_j + \frac{1}{\vartheta\|h_k\|^2}\right)))$$

**Step 4:** if_3 ($v - u < \varepsilon$) then q $\approx \frac{u+v}{2}$

  Else go to Step 2.

  End if_3

n= n+1

End if_1

If_4 ($\sum_{k=1}^{K} p_{k.min} > P_{max}$)

**Step 1:** User admission

End if_4

---

In this algorithm, two scenarios are considered. If maximizing the energy efficiency is not possible, the user conditions are considered in that user acceptance scenario in the cluster, unlike previous studies. Thus, a user which does not meet the minimum power requirements will not be left out, and generally the proposed algorithm outperforms other algorithms.

In Fig. 2, we compare the performance of the proposed algorithm with the proposed algorithm of [13]. As it is seen in Fig. 2, the proposed algorithm performs better than the SIF method. In this comparison, it is assumed that the total transmission power is 1 watt. For instance, the energy efficiency of the proposed method is 6.84 Mbit/j for a fixed power value of 4dBm, which has been improved about 1.23 Mbit/j compared to the proposed method of [13].

In Fig. 3, we present how the number of antennas affects the value of EE. In this implementation, it is assumed that the constant power of the circuit is 7dBm. Obviously, in Fig. 3, the proposed method outperforms the method which has been proposed in [13]. For instance, the energy efficiency of the proposed method is 4.75 Mbit/j for the case of 64 antennas, which has been improved about 0.99 Mbit/j compared to the proposed method of [13].

If the condition that the sum of the minimum required power of users is less than the total transmitted power is not met, we maximize the number of accepted users in the cluster.

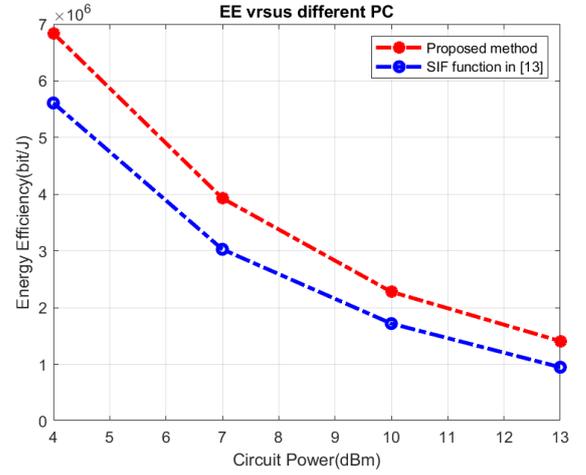

**Fig. 2**. Energy Efficiency versus different $P_c$.

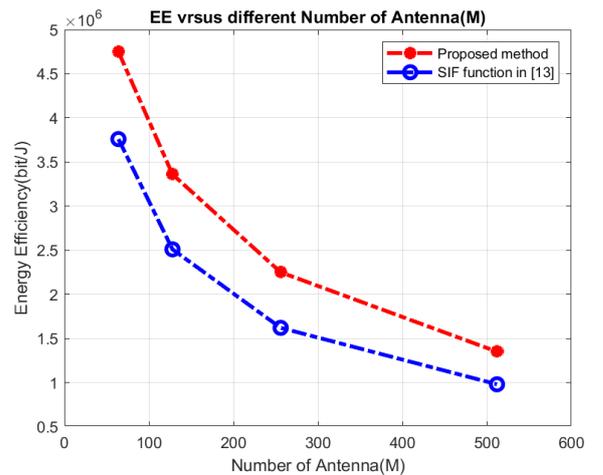

**Fig. 3**. Energy Efficiency versus different $M$.

Fig. 4 shows the average number of admitted users in a cluster. The number of users requesting to be added to the cluster is 9. As it is seen, this figure is plotted for 4 different scenarios, which shows that the average number of users accepted in the cluster increases with the maximum transmission power and the number of requesting users.

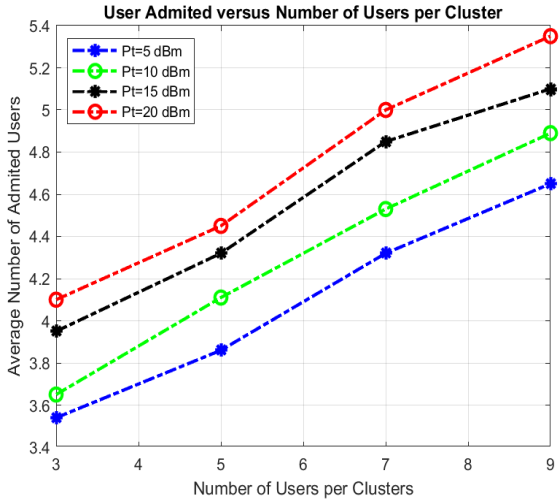

**Fig. 4**. Number of admitted users versus number of users in cluster.

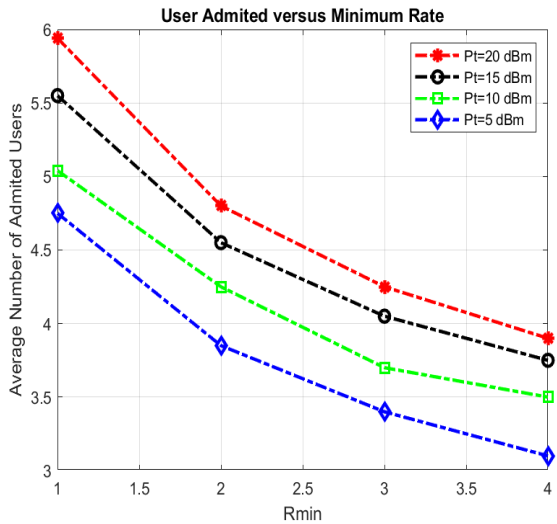

**Fig. 5.** Number of admitted users versus minimum rate value.

Fig. 5 show the performance of the proposed user admission scheme versus minimum rate. Clearly, the average number of users accepted in a cluster decreases with increasing the minimum rate value.
Finally in Table 2, we compare the Energy Efficiency versus the maximum transmission power. According to the results appeared in this table, the proposed algorithm performs better than the SIF method which is proposed in [13].

**Table 2**. Energy Efficiency versus for different values of $P_T$.

| maximum transmission power (w) | EE convergence propose algorithm (Mbit/J) | EE convergence algorithm in [13] (Mbit/J) | Amount of improvement (Mbit/J) |
|---|---|---|---|
| 2 | 4.6251 | 4.2127 | 0.4124 |
| 3 | 4.6027 | 4.2096 | 0.3931 |
| 4 | 4.5869 | 4.2021 | 0.3848 |

## 5. Conclusion

In this paper, we formulated the energy efficient power allocation problem for the massive MIMO system. In addition to the total transmission power, we also considered the minimum power required by users to ensure the quality of service for each user. By comparing the sum of the minimum power required by users and the total transmitted power, the type of optimization problem was determined. According to the presented results, we show that the proposed algorithm performs better than other similar power allocation algorithms.